\documentclass{aa}  
\usepackage{graphicx}
\usepackage{txfonts}
\usepackage{xcolor,soul}

\begin{document}

   \title{Non-synchronous rotations in massive binary systems}

   \subtitle{HD~93343 revisited\fnmsep\thanks{This work is based on observations collected at Complejo Astron\'omico El Leoncito, Las Campanas Observatory, and La Silla Observatory.}}

\titlerunning{Non-synchronous rotation in the HD~93343 binary system}

   \author{C. Putkuri
          \inst{1},
          R. Gamen\inst{1,2}, 
          N.~I. Morrell\inst{3}, 
          S. Sim\'on-D\'iaz\inst{4,5}, 
          R.~H. Barb\'a\inst{6},
          G.~A. Ferrero\inst{1,2},
          J.~I. Arias\inst{6} and
          G.~Solivella\inst{1,2}
          }
 \institute{Instituto de Astrofísica de La Plata (CONICET-UNLP), Paseo del Bosque s/n, 1900, La Plata, Argentina\\
  \email{cputkuri@fcaglp.unlp.edu.ar}
 \and
Facultad de Ciencias Astron\'omicas y Geof\'isicas, Universidad Nacional de La Plata, Paseo del Bosque s/n, 1900, La Plata, Argentina.
\and
Las Campanas Observatory, Carnegie Observatories, Casilla 601, La Serena, Chile
 \and
 Instituto de Astrof\'isica de Canarias, E-38200 La Laguna, Tenerife, Spain
 \and
Departamento de Astrof\'isica, Universidad de La Laguna, E-38205 La Laguna, Tenerife, Spain
  \and
Departamento de F\'isica y Astronom\'ia, Universidad de La Serena, Av. Cisternas 1200 Norte, La Serena, Chile
  }

\authorrunning{Putkuri et al}

   \date{Received ; accepted }

  \abstract
{Most massive stars are in binary or multiple systems. Several massive 
stars have been detected as double-lined spectroscopic binaries and 
among these, the {\it OWN Survey} has detected a non-negligible number
whose components show very different spectral line broadening (i.e., projected rotational
velocities). This fact raises a discussion about the contributing processes, such as
angular-momentum transfer and tidal forces.  }
{We seek to constrain the physical and evolutionary status of one of such systems, the O+O binary \object{HD~93343.}}
{We analyzed a series of high-resolution multiepoch optical spectra to determine the orbital parameters, projected rotational velocities, and evolutionary status of the system.}
{\object{HD~93343} is a binary system comprised of two O7.5~Vz stars that each have minimum masses
of approximately 22~$M_{\odot}$  in a wide and eccentric orbit ($e$ = 0.398~$\pm$0.004; 
$P$=50.432~$\pm$0.001 d). Both stars have very similar stellar parameters, and hence 
ages. As expected from the qualitative appearance of the combined spectrum of the 
system, however, these stars have very different projected rotational velocities 
($\sim$65 and $\sim$325 km~s$^{-1}$, respectively).}
{The orbits and stellar parameters obtained for both components seem to indicate 
that their youth and relative separation is enough to discard the effects of mass 
transfer and tidal friction. Thus, non-synchronization should be intrinsic to their 
formation.}

    \maketitle
%------------------------------------------------------------------------
\section{Introduction}
%------------------------------------------------------------------------

Massive stars play a key role in the evolution of Universe. Their high 
luminosities and strong stellar winds 
sweep away interstellar medium and drive the 
chemical evolution of galaxies. Usually, 
these objects are located in star-forming regions, thereby
heating and enriching neighboring gas clouds where new generations of stars form. In spite of their importance, there are still many missing pieces in our knowledge about the formation and evolution of these cosmic engines. This is partly because their numbers are low and they are highly complex objects.

In recent years, it has become clear that a high percentage of massive stars are part of binary or multiple systems
\citep{1998AJ....115..821M,2011IAUS..272..474S,2013ARA&A..51..269D, 2017IAUS..329..110S}. 
The presence of a nearby companion changes the evolutionary path of the individual components in the system \citep{2012Sci...337..444S} by means of non-negligible tidal forces, mass exchange, and even transfer of angular momentum. During the mass transfer process, the primary may also transfer angular momentum to the secondary, which is thereby spun up \citep{2001A&A...369..939W, 2005A&A...435.1013P,2013ApJ...764..166D}. The efficiency of this process is not well understood and
constitutes one of the largest uncertainties in binary evolution.

The initial distribution of rotational velocities of massive O-type stars, which is 
not well known yet \citep{2012ApJ...748...97R}, could be modified by such a scenario. 
\citet{2013A&A...560A..29R,2015A&A...580A..92R} have 
compared the spin distribution of O-type binary 
components and presumably single stars in the 30 Doradus 
region of the LMC. They found these objects to be similar as a whole, 
but some important differences can be clearly distinguished.

The {\it OWN Survey} \citep{2010RMxAC..38...30B,2014RMxAC..44Q.148B,barba2017} is a spectroscopic monitoring program that aims at observing, with high resolution and signal-to-noise ratio (S/N), a sample of Southern O- and WN-type stars. This survey was started in 2005 and, so far, it has collected about 6700 spectra for $\sim$200 stars, comprising a time sampling per target of at least three epochs. This huge database has allowed the identification of more than 100 radial velocity (RV) variable stars, the determination of orbits for about 50 new binary systems (most of them still unpublished), and the discovery of some stars showing line-profile
variability.
 Among the new findings,
we identified some double line spectroscopic binaries presenting two sets of line profiles affected by a different amount of rotational broadening, respectively.

There are other known examples of systems showing different line broadenings.
We note the following examples: (\textit{i}) The components in \object{HD~37366}, which have projected 
rotational velocities of 30 and 100~km~s$^{-1}$ and
are in an eccentric orbit
($e$=0.33) of 31.9~d \citep{2007ApJ...664.1121B}; (\textit{ii})
Plaskett's star (\object{HD~47129}), whose components have projected 
rotations of $\sim$75 and $\sim$300~km~s$^{-1}$ ($P$=14.396~d) as 
measured by \citet{2008A&A...489..713L};
(\textit{iii}) the components Aa and Ab in the hierarchical multiple system 
$\sigma$~Ori~AB, which have velocities of 135 and 35~km~s$^{-1}$ 
\citep[$P=143.198$~d; ][]{2011ApJ...742...55S,2015ApJ...799..169S}; and 
(\textit{iv}) those in \object{HD~101131}, which  have velocities of 102 and 164~km~s$^{-1}$ in 
a orbital period of 9.65~d \citep{2002ApJ...574..957G}.
Double-lined spectra found in the {\it OWN Survey} would be due to 
a binary system, but this hypothesis has to be tested with RV analysis. 
If confirmed, the different line broadening could be related 
to asynchronous rotation or  different inclinations of the rotation 
axis of each component; this, in turn, could be related to the origin of the binary or its evolution \citep{1981A&A....99..126H}.

This work is the first in a series of papers in which we seek to characterize stars showing such composite spectra.
We explore the possibility of dealing with binary systems whose components are asynchronous rotators, starting with HD~93343.

HD~93343 (CPD --59 2633; $RA_{2000}$=10:45:12.2; $DEC_{2000}$=$-59$:45:00.4; $V$=9.6 mag) 
is an O-type double-lined spectroscopic binary star that is member of the young cluster Trumpler~16.
\citet{1982ApJS...48..145W} determined a spectral type of O7 V(n) and noted the presence of spectral features belonging to a secondary component by first time. Although its RVs have been widely studied by many researchers for decades (e.g., \citet{1991ApJS...75..869L} and 
\citet{1998larm.confE..28S}),
the SB2 nature of HD~93343
was only confirmed  by \citet{2009MNRAS.398.1582R}. These authors inferred spectral types O7-8.5 and O8 for the primary and secondary components, respectively, but were not able to establish an orbital solution. Their data also revealed that the secondary component has broadened lines. \citet{2014ApJS..211...10S} did not detect double lines in their low-resolution data but classified the composite spectrum as O8 Vz. \citet{2016ApJS..224....4M} reclassified the star, removing the qualifier "z", but retaining the O8 V spectral type.
\citet{2016BAAA...58..159P} published a preliminary orbital solution for both components and obtained a mass ratio of 0.87$\pm0.03$.

We present the complete analysis of a dataset comprising 
high-resolution spectra of HD\,93343 (see Sec.~\ref{section2}) spanning more than
ten years.
This paper then presents the individual spectra obtained via spectral disentangling and RV measurements in Sec. \ref{section3}.
In Sec.~\ref{section5} and Sec.~\ref{section4} we present the determination of the orbital solution and  spectroscopic analysis, including spectral classification and determination of the stellar parameters of each component. Finally, we perform an analysis of the evolutionary status of the stars in Sec.~\ref{section40} and discuss the results and their conclusions in Sec.~\ref{section6} and Sec.~\ref{conclusiones}.
%------------------------------------------
 \begin{table*}[!t] 
\caption{Technical details of the instrumental configurations.}            
\label{tabla1}     
\centering                          
\begin{tabular}{c c c c c c c}   
\hline\hline              
Spectrograph & Obser.& Time span & Spectral & $R$ & Reciprocal    &$n$ \\
             &       &           & coverage &     & dispersion    & \\
             &       &           & [\AA]    &     &[\AA~px$^{-1}$]&\\
\hline
\'echelle-REOSC& CAS   & 1994--2016 &3600-6100& 15\,000 & 0.19 &13\\
         \'echelle & LCO   & 2006--2017 &3500-9850& 40\,000 & 0.05 &20\\
           FEROS   & ESO   & 2007--2015 &3570-9210& 46\,000 & 0.03 & 7\\
\hline                              
\end{tabular}
\end{table*}
%------------------------------------------
\section{Observations}\label{section2}
%------------------------------------------
Our observational dataset consists of 40 high-resolution spectra acquired between
1994 and 2017.
We employed the 2.15~m J. Sahade telescope at Complejo Astron\'omico El Leoncito (CASLEO)\footnote{Complejo Astronómico El Leoncito is operated under agreement between the Consejo
Nacional de Investigaciones Científicas y Técnicas de la República Argentina and the
National Universities of La Plata, Córdoba and San Juan.}, 
Argentina; the 2.5~m Ir\'en\'ee du Pont telescope at Las 
Campanas Observatory (LCO), Chile; and the MPG/ESO 2.2~m 
telescope at La Silla Observatory, Chile. For each of these, 
we used their \'echelle spectrographs.
First spectra, at CASLEO, were observed in the context of an international campaign called X--Mega \citep{1999RMxAC...8..131C}.
See Table~\ref{tabla1} for details of the instrumental configuration used in this work, where in successive columns we give the name of the spectrograph, observatory, time span of each dataset, spectral coverage, resolving power ($R$), reciprocal dispersion, and number of obtained spectra ($n$).

At CASLEO and LCO, comparison lamp spectra of Th-Ar were observed immediately
after or before each target integration at the same telescope position. All the spectra were extracted and normalized using the standard IRAF\footnote{IRAF is distributed by the National Optical Astronomy Observatories, which are operated by the Association of Universities for Research in Astronomy, Inc., under cooperative agreement with the National Science Foundation.} routines. The FEROS data were reduced using the standard reduction pipeline provided by ESO. 
%----------------------------------------------------------------------------
\section{Spectral disentangling and radial velocity measurements} \label{section3}
%----------------------------------------------------------------------------
\begin{figure}[!t] 
\resizebox{\hsize}{!}{\includegraphics[angle=-90]{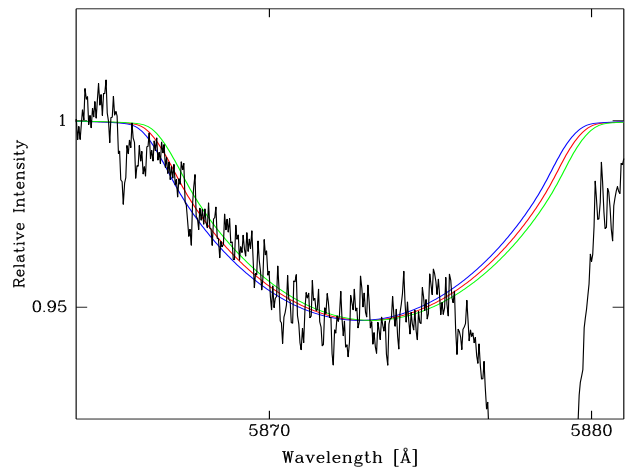}}
 \caption{Comparison of observed composite spectrum (in black) with three
  synthetic templates shifted with different RVs around He {\sc i} $\lambda5875$. 
  The blue line is shifted by 130~km~s$^{-1}$, red by 140, and green by 150\,km~s$^{-1}$. The green curve best fits the spectrum, but the others are not so bad.}
  \label{secundaria}
  \end{figure}
  
To determine the orbital solution of HD~93343, we measured the RVs of both components in the system. For this purpose, we firstly proceeded to disentangle both spectra.
We employed a procedure for separating composite spectra based on the code published by \citet{2006A&A...448..283G}. It basically consists in combining the composite spectra shifted to the RV of one of the components to dilute the spectral features of the other (e.g., secondary). The spectrum combined in this way contains the features of the primary and those of the secondary diluted (as they are at different wavelengths). Then, this template (almost pure primary component) 
is subtracted from the original composite spectra (shifted to the same RV) to 
obtain the spectrum of the other component. These are then 
combined and shifted to the secondary RV.
An initial estimation of the RVs of both components is needed to start the method.
Once the initial templates are generated, the method determines the RVs by means of cross-correlation between each template and subtraction of the other template from the composite spectra. It is assumed that these new RVs are more accurate than the initial RVs, and new templates are generated using the new RVs to shift the spectra. This time, the template of the primary is generated subtracting the template of the secondary. After several iterations, the templates tend to represent the pure spectrum of each component

Because of the peculiarities of this system, we decided to consider some variations to the method.
Regarding the initial RVs, as the secondary component is very broad and it is very difficult to identify the barycenter of the line, we had to be very careful in this task. We determined the central wavelengths of the spectral lines using the {\it{ngauss}} task of IRAF.
The cross-correlation was applied in four different wavelength domains, i.e., 
4461--4478~\AA, 4536--4549~\AA, 4675--4695~\AA, and 5863--5887~\AA, which include the He{\sc i} $\lambda\lambda$4471, 5876 and He{\sc ii} $\lambda\lambda$4542, 
4686 absorption lines.
The RVs of the four lines of the primary component gave similar results (maximum differences $\sim$10~km~s$^{-1}$). These RVs were then averaged to represent the RV of the primary component. 
For the secondary, the differences among RVs were significantly higher and reached values as high as 60~km~s$^{-1}$ in a few cases. 
As a consequence, to represent the secondary motion we used the He{\sc i}~$\lambda$5876 line, which certainly appears to be the least affected by pair 
blending.

The template of the secondary obtained via this raw procedure gave very noisy and 
almost useless results.
We constructed an initial template of the secondary to improve it.
We subtracted a synthetic template of the primary component{\bf \footnote{We used as synthetic template the best-fitting FASTWIND model resulting from the quantitative spectroscopic analysis (see Sect.~\ref{section4}).}} to the best 
composite spectrum obtained during quadratures. In this way, we obtained a template of the secondary component that was used in the first iteration of the disentangling method. 
 
\begin{figure}[!t]
\resizebox{\hsize}{!}{\includegraphics[angle=-90]{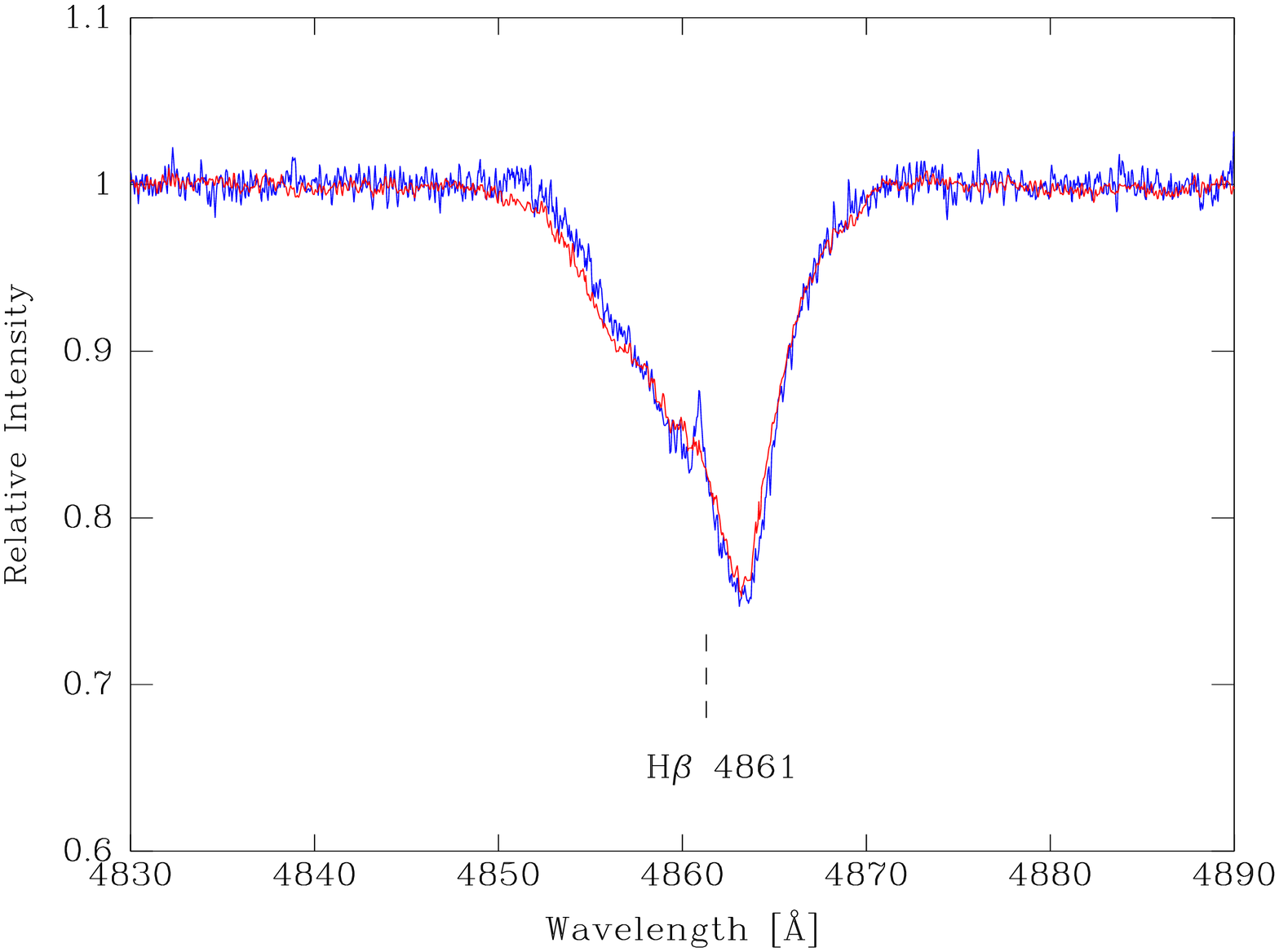}}
\resizebox{\hsize}{!}{\includegraphics[angle=-90]{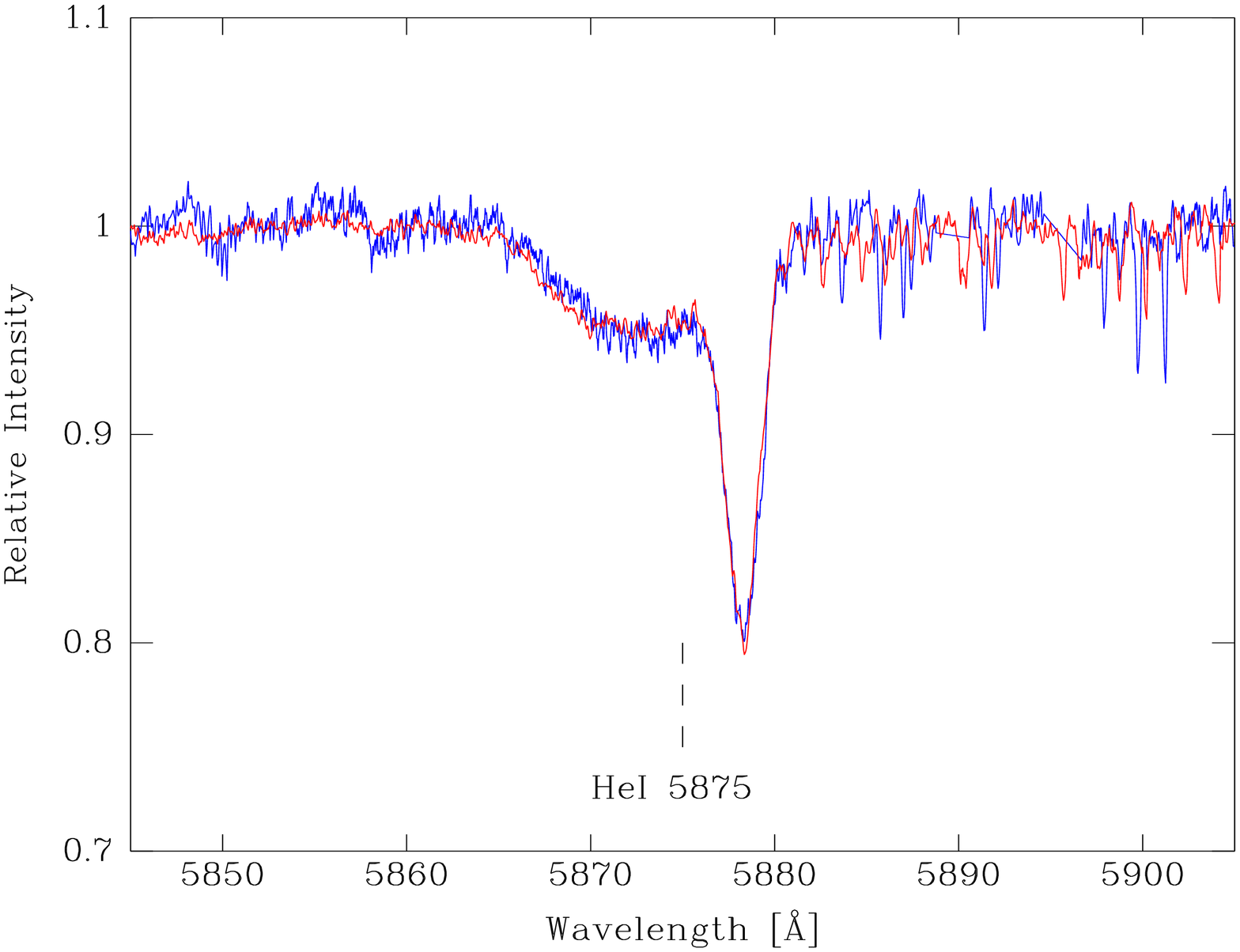}}
  \caption{Comparison between observed composite spectra in phase of
  maximum separation (blue) and the sum of templates shifted with the corresponding 
  RVs of Table~\ref{RVs} (red). 
  Top: around H$\beta$; bottom: around He {\sc i} $\lambda5875$.}
  \label{dissen}
  \end{figure}

To test the confidence of the method, we constructed composite spectra adding both templates, shifted by the obtained RVs and compared these with the observed spectra. We noted that the secondary spectrum did not fit well. 
Actually, fitting the broadened features is an issue that we try to illustrate in
Fig.~\ref{secundaria}. There, we plot synthetic spectra, shifted to three different 
velocities, over an observed spectrum. It can be noted that the differences are subtle, thus errors in RVs could be larger than estimated.
Then, we refined the secondary RVs adopting those that minimize the differences, i.e.,
we always checked the RVs obtained by comparing the original composite spectrum with that constructed from the sum of templates (an example of this is shown in Fig.~\ref{dissen}).
The final RVs used in the orbital solution are showed in Table \ref{RVs}.

\begin{table}
\setlength{\tabcolsep}{1.0mm}
\centering
\caption{Radial velocity measurements of both components in the HD~93343 system. The orbital phase was obtained with the orbital solution given in Table~\ref{orbita}.}
\begin{tabular}{l c ccccc c}
\hline\hline\noalign{\smallskip}
\multicolumn{1}{c}{HJD}& {Phase}& \multicolumn{5}{c}{Radial velocity}&Obser.\\
\multicolumn{1}{c}{[d]}& $\phi$ & & primary &&secondary&\\
\hline\noalign{\smallskip}
\!\!    2449428.758     &       0.22    &         &     59      &       &               &    &    CAS      \\
\!\!    2449429.817     &       0.24    &         &     40      &       &               &    &    CAS      \\
\!\!    2449431.792     &       0.28    &         &     30      &       &               &    &    CAS      \\
\!\!    2449432.739     &       0.30    &         &     15      &       &               &    &    CAS      \\
\!\!    2449433.756     &       0.32    &         &     11      &       &               &    &    CAS      \\
\!\!    2450845.748     &       0.32    &         &     12      &       &               &    &    CAS      \\
\!\!    2452738.606     &       0.85    &         &     -63     &       &               &    &    CAS      \\
\!\!    2453772.660     &       0.35    &         &     -3      &       &       -3      &    &    LCO      \\
\!\!    2453873.610     &       0.35    &         &     -7      &       &       7       &    &    LCO      \\
\!\!    2453876.607     &       0.41    &         &     -25     &       &       35      &    &    LCO      \\
\!\!    2454198.625     &       0.80    &         &     -76     &       &               &    &    LCO      \\
\!\!    2454199.617     &       0.82    &         &     -70     &       &       72      &    &    LCO      \\
\!\!    2454200.654     &       0.84    &         &     -70     &       &               &    &    LCO      \\
\!\!    2454209.682     &       0.02    &         &     125     &       &               &    &    ESO      \\
\!\!    2454246.620     &       0.75    &         &     -79     &       &       80      &    &    ESO      \\
\!\!    2454247.607     &       0.77    &         &     -78     &       &       80      &    &    ESO      \\
\!\!    2454248.530     &       0.79    &         &     -77     &       &               &    &    ESO      \\
\!\!    2454257.606     &       0.97    &         &     64      &       &       -64     &    &    LCO      \\
\!\!    2454258.546     &       0.99    &         &     84      &       &               &    &    LCO      \\
\!\!    2454259.566     &       0.01    &         &     115     &       &       -125    &    &    LCO      \\
\!\!    2454608.539     &       0.93    &         &     5       &       &               &    &    CAS      \\
\!\!    2454609.537     &       0.95    &         &     29      &       &               &    &    CAS      \\
\!\!    2454625.543     &       0.26    &         &     33      &       &       -25     &    &    ESO      \\
\!\!    2454842.849     &       0.57    &         &     -64     &       &               &    &    CAS      \\
\!\!    2454960.621     &       0.91    &         &     -22     &       &               &    &    LCO      \\
\!\!    2454961.639     &       0.93    &         &     0       &       &               &    &    LCO      \\
\!\!    2454962.497     &       0.94    &         &     20      &       &               &    &    LCO      \\
\!\!    2454963.617     &       0.97    &         &     57      &       &               &    &    LCO      \\
\!\!    2454964.551     &       0.98    &         &     86      &       &               &    &    LCO      \\
\!\!    2455976.777     &       0.06    &         &     138     &       &       -140    &    &    LCO      \\
\!\!    2456078.537     &       0.07    &         &     137     &       &       -140    &    &    LCO      \\
\!\!    2457092.648     &       0.18    &         &     73      &       &               &    &    CAS      \\
\!\!    2457094.623     &       0.22    &         &     55      &       &               &    &    CAS      \\
\!\!    2457114.695     &       0.62    &         &     -65     &       &       70      &    &    ESO      \\
\!\!    2457117.616     &       0.68    &         &     -74     &       &       83      &    &    ESO      \\
\!\!    2457557.577     &       0.40    &         &     -22     &       &               &    &    CAS      \\
\!\!    2457591.541     &       0.07    &         &     139     &       &       -140    &    &    LCO      \\
\!\!    2457593.457     &       0.11    &         &     121     &       &       -120    &    &    LCO      \\
\!\!    2457767.775     &       0.57    &         &     -62     &       &               &    &    LCO      \\
\!\!    2457767.799     &       0.57    &         &     -61     &       &       68      &    &    LCO      \\
\hline 
\end{tabular}
\label{RVs}
\end{table}

%--------------------------------------------------------------------
\section{Orbital solution} \label{section5}
%--------------------------------------------------------------------  
\begin{figure}[!t]
  \resizebox{\hsize}{!}{\includegraphics{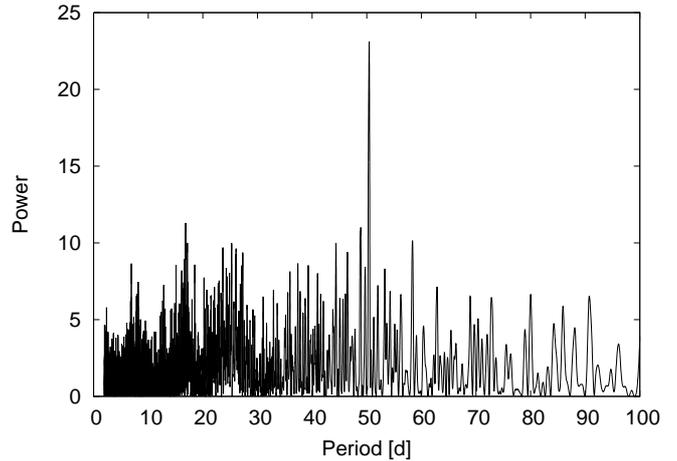}}
  \caption{Periodogram of the primary RVs obtained with NASA Exoplanet Archive Periodogram Service.}
  \label{periodogram}
\end{figure}

We searched for periodicities in our data by means of the NASA Exoplanet Archive Periodogram Service. We employed the Lomb-Scargle \citep{1982ApJ...263..835S} method, obtaining a most probable period ($P$) of 50.45$\pm$0.15~d. In Fig.~\ref{periodogram} we show the strength (named as power) of the candidate periods.

\begin{figure}[!t]
 \resizebox{\hsize}{!}{\includegraphics{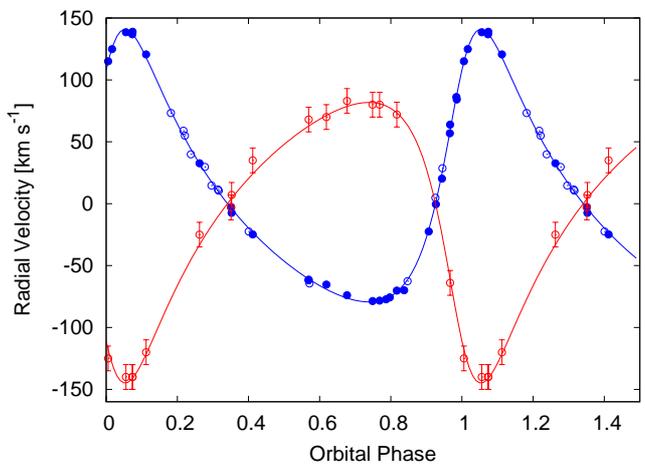}}
  \caption{Radial velocity curves of primary (blue) and secondary (red) components of the binary system HD~93343, calculated with the parameters of the orbital solution shown in Table~\ref{orbita}. Primary RV errors are smaller than sizes of the symbols. Primary RVs  were weighted according the resolution of their spectra, LCO and La Silla data with 1 (filled circles), and CASLEO with 0.5 (open circles). For the secondary all data were weighted with 0.5 (open circles).}
  \label{RVcurves}
\end{figure}

With this $P$ as initial value, we determined the orbital solution using the {\sc gbart} code\footnote{Based on the algorithm of \citet{1969RA......8....1B} and implemented by F. Bareilles (available at \url{http://www.iar.unlp.edu.ar/~fede/pub/gbart}).}, letting all the parameters free. 
The RVs of CASLEO were weighted by half owing to its lower resolution respect to 
the LCO and FEROS values. 
The orbital parameters obtained are given in Table~\ref{orbita} and the RV curves are depicted in Fig.~\ref{RVcurves}.
It is important to highlight that the RV amplitude of the secondary component is lower than previously reported by \citet{2016BAAA...58..159P}, which results in lower minimum masses.

\begin{table}[!t]
\setlength{\tabcolsep}{1.0mm}
\centering
\caption{Orbital solution of HD~93343.}
\begin{tabular}{l rcl c rcl}
\hline\hline\noalign{\smallskip}
Parameter& \multicolumn{3}{c}{Primary}&& \multicolumn{3}{c}{Secondary}\\
\hline\noalign{\smallskip}
\!\!$P$ [d]                               &\multicolumn{3}{r}{50.432}&$\pm$&\multicolumn{3}{l}{0.001}  \\
\!\!$T_{\rm periastron}$ [HJD]     &\multicolumn{3}{r}{2\,455\,973.97}&$\pm$&\multicolumn{3}{l}{0.05}    \\
\!\!$V_{0}$ [km s$^{-1}$]               &\multicolumn{3}{r}{$0.06$}&$\pm$&\multicolumn{3}{l}{0.31} \\
\!\!$e$                                 &\multicolumn{3}{r}{0.398} &$\pm$&\multicolumn{3}{l}{0.004}  \\
\!\!$\omega$ [deg]                      &\multicolumn{3}{r}{-45.5} &$\pm$&\multicolumn{3}{l}{0.6} \\
\!\!$K_{i}$ [km s$^{-1}$]               & 110.1         &$\pm$&  0.5 && 113.3          &$\pm$& 0.8\\
\!\!$a_{i}$  $\sin i$  [R$_{\odot}$]    & 100.7       &$\pm$& 0.7  && 103.6       &$\pm$& 0.7\\
\!\!$M_{i}$  $\sin^{3} i$  [M$_{\odot}$]& 22.8       &$\pm$& 0.9 && 22.1       &$\pm$& 0.7\\
\!\!$q$ [M$_{2}$/M$_{1}$]               &\multicolumn{3}{r}{0.97}&$\pm$&\multicolumn{3}{l}{0.01}   \\
\hline 
\end{tabular}
\label{orbita}
\end{table}

%----------------------------------------------------------------------------
%-----------------------------------------------------------------------------
\section{Spectroscopic analysis} \label{section4}
%----------------------------------------------------------------------------
%----------------------------------------------------------------------------
  
\subsection{Spectral classification} \label{sec4.1}
 
We used the templates resulting from the disentangling method to provide a spectral classification of both components of the binary system. To this aim, we degraded the templates to a resolving power ($R$) of 2500 and compared these with the set of spectra of O-type standards proposed by \citet{2011ApJS..193...24S} and \citet{2016ApJS..224....4M}.
In addition, we followed the guidelines for spectral classification of this type of objects indicated in
those works.

\begin{figure*}[!t]
\resizebox{\hsize}{!}{\includegraphics[angle=-90]{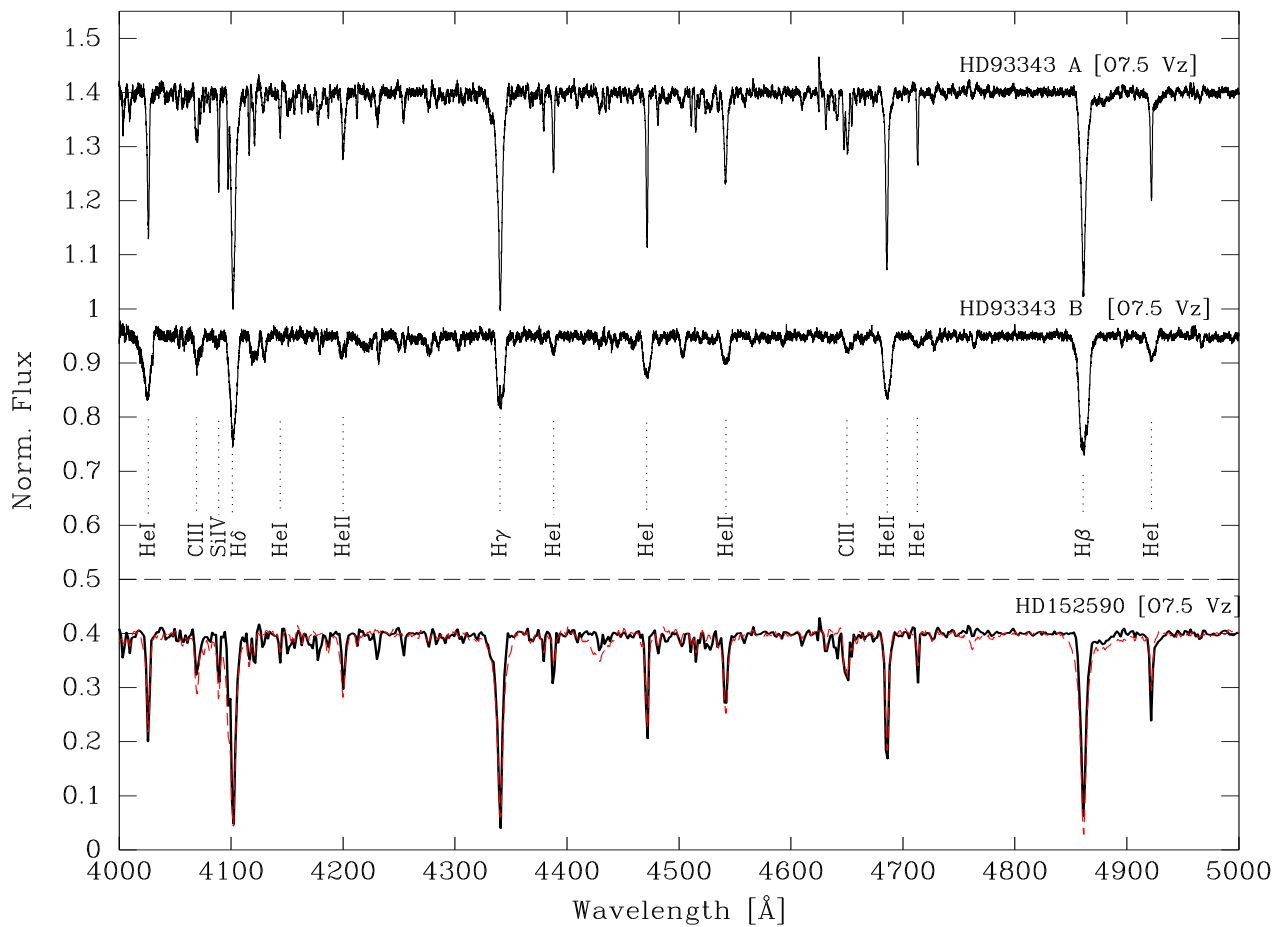}}
  \caption{High-resolution disentangled spectra of both components of the binary system HD~93343 (first two solid black lines). Features that are relevant for spectral classification are indicated. Some artifacts appear on the spectrum of the secondary component, mostly in the wings of hydrogen lines that originate in the disentangling process.
  In the lower box we present the primary component degraded in resolution and overplotted with the O7.5~Vz standard star from \citet[][red dashed line]{2016ApJS..224....4M}.}
  \label{spectra}
  \end{figure*}

One of the major criteria for spectral type is the ratio He {\sc ii} $\lambda$4542/He {\sc i} $\lambda$4471. In both spectra, this ratio resulted in less than unity, thus indicating spectral types later than O7. The ratios He {\sc ii} $\lambda$4542/He {\sc i} $\lambda$4388 $>$ 1  and He {\sc ii} $\lambda$4200/He {\sc i} $\lambda$4144 $\gg$ 1 are in agreement with an O7.5-8 subtype.
On the other hand, the ratio He {\sc ii} $\lambda$4686/He {\sc i} $\lambda$4713 $\gg 1$ led to a luminosity class V. 
The unusually strong absorption in the He {\sc ii} $\lambda$4686 line observed in both templates indicates the addition of the "z" qualifier. The agreement of the template with the O7.5 Vz standard from \citet{2016ApJS..224....4M} is evident for the primary (see Fig.~\ref{spectra}) and, because of the high rotational broadening, less clear for the secondary. However the Vz classification for both components is confirmed through the  quantitative criterion proposed by \citet{2016AJ....152...31A}, which states that the ratio of the equivalent width of the He {\sc ii} $\lambda$4686 line to the maximum between the equivalent widths of the He {\sc i} $\lambda$4471 and He {\sc ii} $\lambda$4542 must be greater or equal than 1.1.

Both stars present overall similar spectra as can be seen in Fig.~\ref{spectra}, but it can be noticed that C {\sc iii} $\lambda$4070
presents stronger in the secondary component than in the primary. This could indicate a surface
chemical enrichment due to fast rotation, since this favors the transport
of heavier elements synthesized in the nucleus toward the outer stellar layers \citep{2000A&A...361..101M}.

%----------------------------------------------------------------------------
\subsection{Quantitative spectroscopic analysis} \label{sec4.2}
%----------------------------------------------------------------------------

We employed the \textsc{iacob-broad} tool \citep{2014A&A...562A.135S} to obtain estimates for the projected rotational velocity ($v$\,sin\,$i$) and the macroturbulence broadening ($v_{\rm mac}$) of each component of the binary system. To this aim, we used the O\,{\sc iii}\,$\lambda$5592 line present in each of the corresponding disentangled spectra (this time not degraded in resolution). As expected from the qualitative appearance of the original  composite spectra (see Fig.~\ref{dissen}), the derived values of $v$\,sin\,$i$ are very different for the narrow and broad components ($\sim$65 and $\sim$325 km\,s$^{-1}$, respectively; see also Table~\ref{tablagbat}).

Subsequently, we proceeded with the determination of other spectroscopic parameters such as the effective temperature ($T_{\rm eff}$) and stellar gravity (log\,$g$). To this aim, we first tried to perform a quantitative spectroscopic analysis of the individual disentangled spectra with \textsc{iacob-gbat}\footnote{The IACOB grid-based algorithm tool (\textsc{iacob-gbat}) is a user-friendly IDL package based on standard techniques for the quantitative spectroscopic analysis of O stars, which has been automated by applying a $\chi^2$ algorithm to a large grid of synthetic spectra computed with the FASTWIND stellar atmosphere code (Santolaya-Rey et al. 1997; Puls et al. 2005; Rivero Gonzalez et al. 2012a). Recent applications of this tool can be found in Sab\'in-Sanjulian et al. 2014, 2017 and Holgado et al. 2018.} \citep{2011JPhCS.328a2021S}. However, we found that the solution provided by \textsc{iacob-gbat} was not completely satisfactory, especially in the case of the secondary star, because of the presence of some spurious artifacts affecting the wings of the hydrogen lines (see Fig.~\ref{spectra}) introduced during the disentangling process. We hence decided to follow a different approach; namely, the stellar parameters of the two components were obtained directly and simultaneously from the analysis of one of the original spectra. In particular, we considered one of the FEROS spectra with largest separation between lines.

The combined synthetic spectra to be fitted to the observed spectrum were constructed
using spectra from the grid of FASTWIND models with solar metallicity 
included in \textsc{iacob-gbat}. The spectrum of each component was 
convolved to the corresponding $v$\,sin\,$i$ and $v_{\rm mac}$, 
shifted in RV, and scaled by a certain factor $d_i$, where $\sum d_i$=1. Then, 
the two synthetic spectra were added together, and the combined spectrum 
overplotted to the observed spectrum.

During the analysis process, we fixed the associated helium abundances
($Y_{\rm He}$), microturbulent velocities ($\xi_{\rm t}$), and the 
two wind parameters considered in the grid of 
FASTWIND models \citep[$\beta$ and log\,$Q$; see][]{2011JPhCS.328a2021S} 
to characteristic values commonly obtained 
for Galactic mid O-type dwarfs (basically 
$Y_{\rm He}$\,=\,0.10, $\xi_{\rm t}$\,=\,10~km\,s$^{-1}$, $\beta$\,=\,0.8, 
log\,$Q$\,=\,--13.5). Also, given that the orbital solution indicates 
that the mass ratio of the stars is close to unity, we assumed 
that both components of the binary system contribute 50\,\% to the 
global spectrum (i.e., $d_i$\,=\,0.5). As a result, only the effective 
temperatures and gravities remained as free parameters to be determined in 
the analysis.

\begin{figure*}[!t] 
\resizebox{\hsize}{!}
{\includegraphics[angle=90]{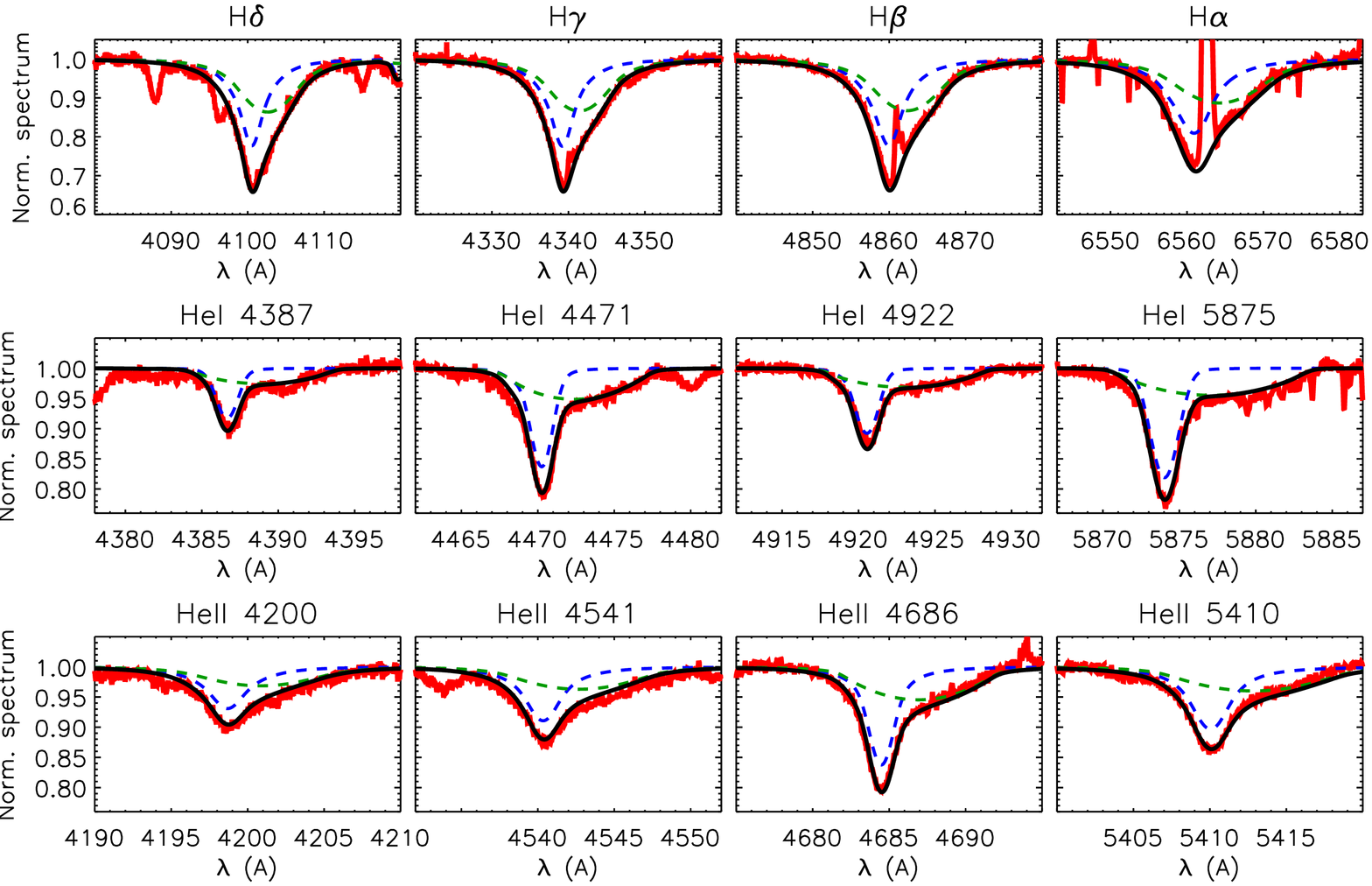}}
  \caption{
Analysis by FASTWIND of one of the FEROS combined spectra of 
HD\,93343 with largest separation between the lines of both components. Solid red and black lines correspond to the observed spectrum and best combined synthetic spectrum, respectively. The individual best-fit synthetic spectra for each component are overplotted with green and blue dashed lines, respectively.}
  \label{figcomparasion}
\end{figure*}

The best-fit solution was obtained by visual comparison of 
the original and combined synthetic spectra around the 
H and He\,\textsc{i-ii} lines, which are commonly assumed as 
a diagnosis for the determination of stellar parameters of O 
stars (e.g., \citet{1992A&A...261..209H,2002A&A...396..949H}; 
\citet{2004A&A...415..349R}). The final fit is shown in Fig.~\ref{figcomparasion}, 
where the relative contribution of each component to the global 
spectrum is also presented. This figure serves to illustrate the quality 
of the best-fit solution as well as the difficulty in determining precisely 
the gravity of the two components from the wings of the hydrogen Balmer lines. 
We note that given the complexity of the analysis and the similarity of both stars, 
except for the case of the associated projected rotational velocities, 
we basically obtained the same $T_{\rm eff}$ and log\,$g$ for 
the primary and secondary components of the system. The resulting values
are summarized in Table~\ref{tablagbat}.

To calculate the corresponding radii, luminosities, and spectroscopic
masses, the absolute magnitudes of each component are required. On the one hand, we could assume the value provided in the calibration by \citet{2005A&A...436.1049M} for an O7.5~V star ($M_V~=~-~4.5$).  
On the other hand, the individual absolute magnitudes can be estimated if the distance, apparent magnitude ($m_V$ diluted), and amount of extinction ($A_{V}$) are known. 
We gleaned these parameters in the available literature 
and adopted a distance of 2.7~kpc \citep{2016A&A...595A...1G,2018arXiv180409365G}. We determined
$m_V$ and color excess from observational values cited in the Simbad Astronomical Database \citep{2000A&AS..143....9W} and intrinsic colors from \citet{1994MNRAS.270..229W}.  
To obtain the absorption $A_V$, we assumed $R_V=4.4\pm0.2$ from \citet{2012AJ....143...41H}.
Thus, we obtained $E(B-V)=0.53$ and $A_{V}=2.33$.
Using this information and assuming that both stars equally contribute to the total stellar flux, we ended up with an individual absolute magnitude for each component of  $M_V~=~-~4.19$.
As both approaches result in rather different values of $M_{\rm V}$, and for comparative purposes, we keep in Table~\ref{tablagbat} the two associated sets of derived $R$, $L$ and $M_{\rm sp}$. Again, we note that the quoted values are representative of both components since we are basically considering the same $T_{\rm eff}$, log\,$g,$ and $M_{\rm V}$.

   \begin{table}[!t]
\caption{Summary of spectroscopic and physical parameters of the two components of HD~93343 estimated in this work. Apart from the derived $v$~sin~$i$, the rest of parameters are basically the same for both components (see Sect.~\ref{sec4.2}); hence, they are only quoted once.}
\label{tablagbat}      
\centering                          
\begin{tabular}{c c c }    
\hline       \hline    \noalign{\smallskip}   
 Parameter & Primary & Secondary   \\
 \noalign{\smallskip}
\hline  \noalign{\smallskip}   
 
           $v \sin i$ (km\,s$^{-1}$) & 63\,$\pm$\,5 & 325\,$\pm$\,50\\
           $v_{\rm mac}$ (km\,s$^{-1}$) & 48\,$\pm$\,5 & --- \\
           $T_{\rm eff}$ (K)& \multicolumn{2}{c}{36000 $\pm$ 2000} \\
           $\log g$ (dex) & \multicolumn{2}{c}{3.85 $\pm$ 0.15} \\
 \noalign{\smallskip}
\hline  \noalign{\smallskip} 
\multicolumn{3}{c}{Considering $M_{\rm V}=-4.5$ mag}\\
 \noalign{\smallskip}
\hline  \noalign{\smallskip} 

       $R$ (R$_{\odot}$) & \multicolumn{2}{c}{8.7 $\pm$ 0.1}\\

        log ($L/L_{\odot}$) & \multicolumn{2}{c}{5.06 $\pm$ 0.02}\\
        $M_{\rm sp}$ (M$_{\odot}$) & \multicolumn{2}{c}{20.7 $\pm$ 4.6}\\

 \noalign{\smallskip}
\hline  \noalign{\smallskip} 
\multicolumn{3}{c}{Considering $M_{\rm V}=-4.19$ mag}\\
 \noalign{\smallskip}
\hline  \noalign{\smallskip}

$R$ (R$_{\odot}$)       & \multicolumn{2}{c}{7.5 $\pm$ 0.1}\\
log ($L/L_{\odot}$)         & \multicolumn{2}{c}{4.93 $\pm$ 0.02}\\
$M_{\rm sp}$ (M$_{\odot}$)        & \multicolumn{2}{c}{15.8 $\pm$ 4.0}\\

\noalign{\smallskip} 
\hline     
\end{tabular}
\end{table}

%-------------------------------------------------------------------
\section{Evolutionary status} \label{section40}
%--------------------------------------------------------------------  
We used the Bayesian automatic tool {\sc bonnsai} \citep{2014A&A...570A..66S}\footnote{The {\sc bonnsai} web-service is available at {www.astro.uni-bonn.de/stars/bonnsai}.} to 
characterize the evolutionary status of the components in the HD~93343 system.
The {\sc bonnsai} project is a 
statistical method, which matches the entire available observables 
(more specifically $T_{\rm eff}$, $\log g$, $\log L,$ and $v \sin i$)
simultaneously 
to stellar models from \citet{2011A&A...530A.115B}
while taking observed uncertainties and prior knowledge. 
 
\begin{figure}[!t]
  \resizebox{\hsize}{!}{\includegraphics{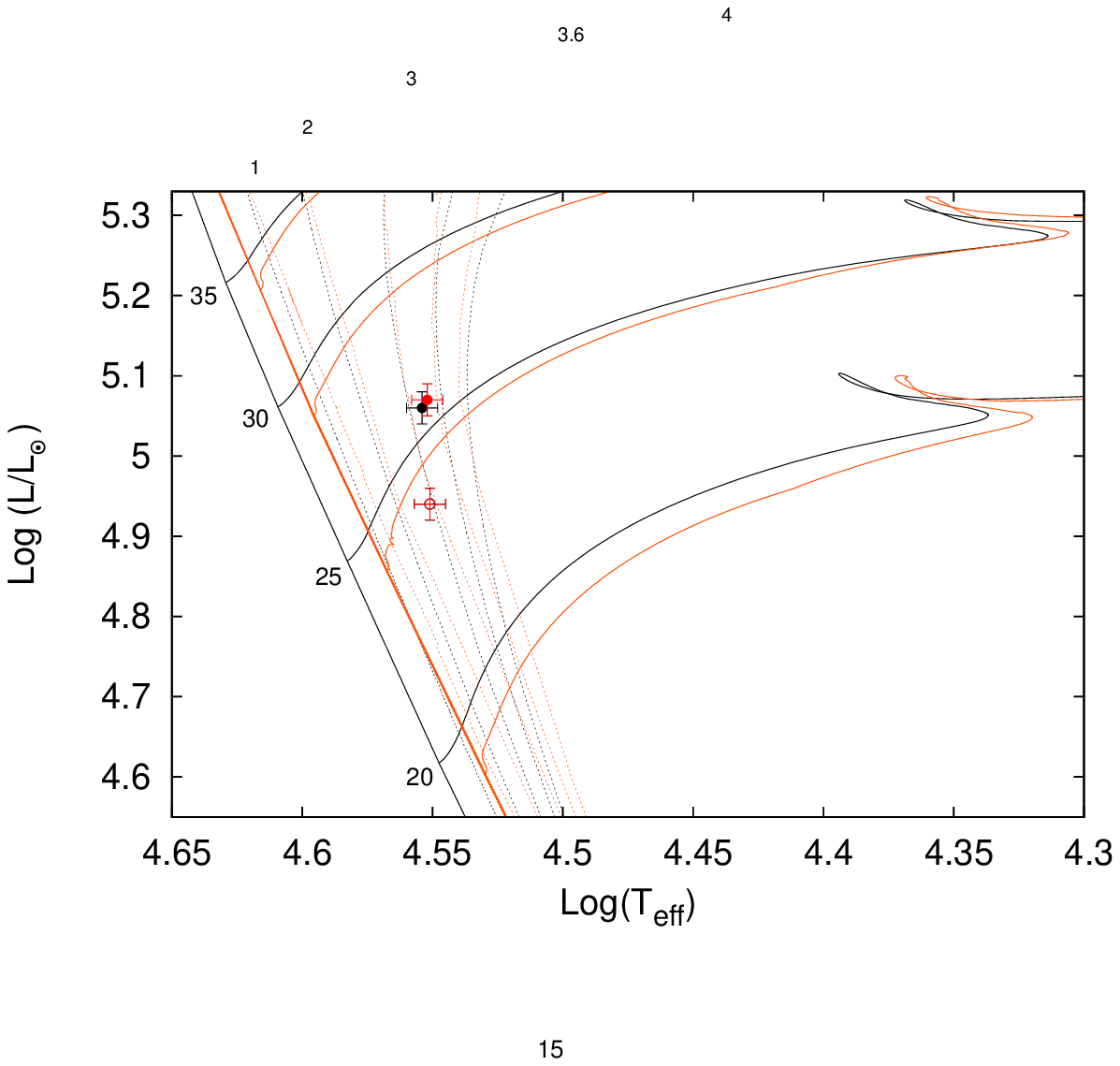}}
  \caption{
Evolutionary tracks in the HR diagram for stellar masses in the range 20--35 M$_{\sun}$ taken from \citet{2011A&A...530A.115B}. The vertical lines correspond to the
isochrones of 1, 2, 3, 3.6, and 4 Myr (dotted lines) and the ZAMS (solid line). Horizontal lines depict the evolutionary tracks: black solid lines identify values of
rotational velocities
in the range 53-61 km~s$^{-1}$(that is, for the primary). Orange, the same for the
secondary (rotation in the range 320-330 km~~s$^{-1}$).
The points represent the location of the HD~93343 binary components
(black for the primary and red for the secondary): filled
and empty dots for $M_{V}$ from \citet{2005A&A...436.1049M} and for $M_{V}$ estimated
from distance modulus calculated in Sec.~\ref{sec4.2}, respectively.}
  \label{HR}
\end{figure}

We run {\sc bonnsai} using the $T_{\rm eff}$, $\log g$, and $v$~sin~$i$ resulting from the quantitative spectroscopic analysis (see Section~\ref{sec4.2}), and the two estimates of $\log L$ quoted in Table~\ref{tablagbat}.
The results are presented in Table~\ref{bonssai}
and depicted 
in a Hertzprung-Russell (HR) diagram in Fig.~\ref{HR}.  
In this figure, we also plot a set of evolutionary tracks and isochrones calculated 
by \citet{2011A&A...530A.115B} with initial rotational velocities in the range 53--61 
km~s$^{-1}$, to represent the primary component, and 
320--330 km~s$^{-1}$ (orange) to depict the secondary.
As can be seen, both stars are located near the zero age main sequence (ZAMS)
between the isochrones of $\sim$3.0 and 3.6 Myr.

%-----------------------------------------------------------
\section{Discussion} \label{section6}
%-----------------------------------------------------------
We made a series of analyses on the disentangled spectra. Firstly, we confirmed the very
different broadening presented by both components, which we visually noted and were also indicated by
\citet{2009MNRAS.398.1582R}. We calculated that the projected rotational velocity of the
secondary is about five times that of the primary star. Hence, its spectral features are more 
diluted in the composite spectrum. However, the spectral type of this star is similar to that of the primary,
even the "z" qualifier is appropriate for both stars. 

We obtained a different orbital solution than the preliminary published
by \citet{2016BAAA...58..159P}. 
This time, we revised the secondary RVs very carefully and employed new ways to measure the RVs,
i.e., adjusting these values by comparison of the composite original spectrum with that created from the
sum of both templates, shifted to the determined RV  (see Sec.~\ref{section3}).
We found a smaller orbit for the broad component and therefore lower minimum masses than before. 
In this work, the determined minimum masses are $M \sin^3 i \approx 22$~M$_{\sun}$ for both components, which are nearly those expected for their spectral types. Thus, the orbital inclination should be close to 90$^\circ$.
We performed some raw trials with the PHysics Of Eclipsing BinariEs ({\sc phoebe} 0.31a version) code \citep{2005ApJ...628..426P} to our RV data and the photometry published by \citet{2009MNRAS.398.1582R} generously provided by E. Fernández Lajús. 
As no evident eclipses are detected in the light curve, we estimated a maximum orbital inclination of 84$^\circ$, which would produce sharp eclipses of about 0.01 magnitudes in the threshold of detectability of those data.
More accurate photometric observations are critical to determine a reliable inclination,
and thus absolute masses of two O7.5~Vz-type stars.

We performed the analysis of the physical properties and evolutionary status 
employing the {\sc iacob-gbat} and {\sc bonnsai} tools.
The range of initial evolutionary masses obtained for both stars, i.e., 23--26~M$_{\sun}$, is well within the errors of the dynamical spectroscopic mass ($M \sin^3 i \approx 22$~M$_{\sun}$).
On the other hand, the spectroscopic masses ($\sim$ 16--21~M$_{\sun}$) resulted in values considerably lower than those. This is a long-standing problem known as mass discrepancy \citep{1992A&A...261..209H}. It is tailored discussed in \citet{2018A&A...613A..12M}, where the discrepancy seems to be more evident in the range of masses around 20~M$_{\sun}$ (see their Figure~11). 

It is important to note the role of rotational broadening on the Vz phenomenon. 
\citet{2014A&A...564A..39S} found that higher values of $v \sin i$ 
generally favor the Vz characteristic at relatively low temperatures ($T_{\rm eff} <$ 38\,000~K), they also found that at an intermediate range of temperatures, between 35\,000 K and 
40\,000~K, a star with a modest wind strength also appears as Vz.
Nevertheless, the position of both components of HD~93343 in the HR diagram seems to point out
the youth of the system.
We crudely estimated the timescales for synchronization ($t_{sync}$) and circularization ($t_{circ}$) for the system by means of the expressions given by \citet{1977A&A....57..383Z}. For this, we used the grid of stellar models by \citet{2004A&A...424..919C}, considering stellar masses of $\sim$20~M$_{\sun}$ and extrapolating the term of moment of inertia ($I/MR^{2}$) from the data provided by \citet{1975A&A....41..329Z} in tabular form, as a function of stellar mass.
For our system we find $t_{sync}$~$\sim$~6~$\times$~10$^{8}$ years and $t_{circ}$ $\sim$ 8~$\times$~10$^{11}$ years. The strong dependence of $t_{sync}$ and $t_{circ}$ on the ratio $a/R$ produces, for this wide binary, timescales for synchronization and circularization that are much larger than the estimated age of the system.

Exploring the origin of the asynchronous rotation of both components, we analyzed the mass transfer scenario. 
In order to determine whether or not a star fills its critical Roche lobe, that is, if the stellar radius attains the radius of the sphere with a volume equal to that of the Roche lobe ($R_{st}$ $=$ $R_{L}$), 
we calculated the Roche lobe radii of the system during the periastron passage. This can be approximated by Eq.(45) of \citet{2007ApJ...660.1624S}, i.e., $R_{\rm L1,peri}^{\rm Egg}=a(1-e)(0.49q^{2/3})/(0.6 q^{2/3}+\ln(1+q^{1/3}))$.
We found $R_{\rm L1,peri}^{\rm Egg}$ = 21.5~R$_{\sun}$ and $R_{\rm L1,peri}^{\rm Egg}$ = 21.8~R$_{\sun}$ for the primary and secondary, respectively. These results imply that the stellar radii are well inside ($\sim$~40$\%$) the respective Roche lobes.
Hence, mass and angular momentum exchange 
should not have started yet and both stars are evolving independently, i.e., following single-star evolution so far. 

The idea  
that in close binaries the rapid rotation is reached when the mass transfer is accompanied by angular momentum exchange \citep{1981A&A...102...17P} is not applicable here. 
The scenario of HD~93343 is very different: 
it is a young, eccentric, and wide binary that has components whose 
radii occupy little volume of their respective Roche lobes 
and whose projected rotational velocities are considerably different.
Thus, the fact that this system is asynchronous should be related to its formation ambient.

 \begin{table}[!t]
\setlength{\tabcolsep}{1.0mm}
\centering
\caption{
Evolutionary parameters of the HD~93343 system as determined with {\sc BONNSAI}
 }
 \begin{tabular}{l rcl c rcl l rcl}
\hline\hline\noalign{\smallskip}
Binary & M$_{V}$ (mag) &log $(L_{\rm theo}/L_{\sun}$) & M$_{\rm ini}$(M$_{\sun}$)&  $\tau$ (Myr) \\

\hline\noalign{\smallskip}
\!\!HD~93343~A  &  $-$4.50  &  5.07$^{+0.02}_{-0.02}$  &26.00$^{+0.90}_{-1.09}$ &  3.50$^{+1.03}_{-0.67}$ \\
\\
\!\!HD~93343~A  &  $-$4.19  &  4.93$^{+0.02}_{-0.02}$  &23.40$^{+0.50}_{-0.44}$ &  3.02$^{+0.47}_{-0.49}$ \\
\\
\!\!HD~93343~B  &   $-$4.50  &  5.06$^{+0.03}_{-0.01}$ &25.80$^{+0.60}_{-0.71}$ & 3.24$^{+0.39}_{-0.40}$ \\
\\
\!\!HD~93343~B  &  $-$4.19  &  4.93$^{+0.02}_{-0.02}$  &22.40$^{+0.89}_{-1.30}$ &  4.60$^{+1.82}_{-1.08}$ \\

\hline
\end{tabular}
\label{bonssai}
\end{table}

%--------------------------------------------------------------------
\section{Conclusions} \label{conclusiones}
%-----------------------------------------------------------

We have observed the spectroscopic massive binary HD~93343 obtaining a set of high quality spectra,
which allowed us to apply the disentangling technique of \citet{2006A&A...448..283G} to 
calculate the individual spectra, and classify both of these
as O7.5~Vz. 

We measured the RVs in each obtained spectrum using the individual disentangled spectra
(see Section~\ref{section3}) and determined a new SB2 orbital solution. The system resulted in a wide ($P=50.432\pm0.001$~d) and eccentric ($e=0.398\pm0.004$) orbit.

The stellar physical parameters were obtained directly from the quantitative spectroscopic analysis of one of the original FEROS spectra, where the lines of the stars are best separated (see Sec.~\ref{section40} for
details and Table~\ref{tablagbat} for results). 
The major difference between components is the projected rotational velocities 
of 63 and 330~km~s$^{-1}$ for the narrow and broad component, respectively. 

The dynamical minimum masses obtained agree with the 
spectroscopic masses (computed via the {\sc iacob-gbat} tool), and
the evolutionary masses (estimated by means of {\sc bonnsai}) and are also in agreement with 
the masses 
expected from the observational $T_{\rm eff}$ scales by \citet{2005A&A...436.1049M}.
 
The evolutionary status of the components in the system, i.e., the evolutionary 
masses and ages, was characterized by means the {\sc BONNSAI} tool. An age of 3.4--3.8~Myr was
derived, depending on the adopted $M_V$.
Following \citet{2013A&A...560A..16M}, a 20--30 M$_\odot$ star reaches the terminal age main sequence (TAMS) in about 8--9~Myr and spends 80~\% of this time in the dwarf phase \citep{2017A&A...598A..56M}, hence the system is closer to the ZAMS than the TAMS.
Thus, their comparative youth and the obtained radius
(well inside their respective Roche lobes), indicate that no
star has experienced Roche lobe overflow and an exchange of matter or angular momentum has not occurred yet. 

Combining findings from this study, i.e., the very different projected rotational velocities, their youth, and the wide orbit, we can conclude that the spins of the two components are not synchronized. The current rotational configuration should be intrinsic to the birth of the system.
We will continue our study with the other systems in the sample, looking for possible relations between asynchronism, eccentricity, orbital period, and age.

%__________________________________________________________________

\begin{acknowledgements}
We thank the referee for her/his valuable comments and
suggestions.
CP, RG, NM, GF, and GS are Visiting Astronomers, Complejo Astronómico El Leoncito operated under agreement between the Consejo Nacional de Investigaciones Científicas y Técnicas de la República Argentina and the National Universities of La Plata, Córdoba and San Juan.
We thank the directors and staff of CASLEO, LCO, and La Silla/ESO for support and hospitality during our observing runs.\\

This work has made use of data from the European Space Agency (ESA) mission
{\it Gaia} (\url{https://www.cosmos.esa.int/gaia}), processed by the {\it Gaia}
Data Processing and Analysis Consortium (DPAC,
\url{https://www.cosmos.esa.int/web/gaia/dpac/consortium}). Funding for the DPAC
has been provided by national institutions, in particular the institutions
participating in the {\it Gaia} Multilateral Agreement.\\

S.S.-D. acknowledges financial support from the Spanish Ministry of Economy and Competitiveness (MINECO) through grants AYA2015-68012-C2-1 and Severo Ochoa SEV-2015-0548, and grant ProID2017010115 from the Gobierno de Canarias.

\end{acknowledgements}

\bibliographystyle{aa}
\bibliography{own}

\end{document}